%% file: main.tex
\documentclass[12pt]{elsarticle}
\biboptions{sort&compress}
\pdfoutput=1
\usepackage[numbers]{natbib}
\usepackage{url}
\usepackage[hmargin=2.35cm]{geometry}
\usepackage{caption}
\usepackage{subcaption}
\usepackage{xspace}
\usepackage{graphicx}
\usepackage{epstopdf}
\usepackage{mciteplus}

\usepackage{ifthen} 
\newboolean{uprightparticles}
\setboolean{uprightparticles}{false} 
\newboolean{articletitles}
\setboolean{articletitles}{true} 
\newcommand{\cmark}{\ding{51}}

\journal{EPJ Plus}

\begin{document}
\title{Heavy-quark opportunities and challenges at FCC-ee}
\author[clermont]{St\'ephane Monteil} 
\author[oxford]{Guy Wilkinson}
\address[clermont]{Universit\'e Clermont Auvergne, CNRS/IN2P3, LPC, Clermont-Ferrand, France}
\address[oxford]{Department of Physics,  University of Oxford, Oxford, United Kingdom}

\date{Received: \today / Revised version: \today }
%
\begin{abstract}
The abundant production of beauty and charm hadrons in the $5 \times 10^{12}$ $Z^0$ decays expected at FCC-ee offers outstanding opportunities in flavour physics that in general exceed those available at Belle II, and are complementary to the heavy-flavour programme of the LHC.  A wide range of measurements will be possible in heavy-flavour spectroscopy, rare decays of heavy-flavoured particles and $C\!P$-violation studies, which will benefit from the low-background experimental environment, the high Lorentz boost, and the availability of the full spectrum of hadron species. 
This essay first surveys the important questions in heavy-flavour physics, and assesses the likely theoretical and experimental landscape at the turn-on of FCC-ee.  From this certain measurements are identified where the impact of  FCC-ee will be particularly important. A full exploitation of the heavy-flavour potential of  FCC-ee places specific constraints and challenges on detector design, which in some cases are in tension with those imposed by the other physics goals of the facility.  These requirements and conflicts are discussed.
\end{abstract} 
%
%
\maketitle

\input{introduction}

\input{landscape}

\section{Flavour-physics opportunities at FCC-ee: some key measurements}
\label{section:experiment}

FCC-ee will perform sensitive studies of $C\!P$-violating phenomena that will allow for improved knowledge of the Unitarity Triangle angles $\alpha$, $\beta$ and $\gamma$, and the phase $\phi_s$ between mixing and decay in the $B^0_s-\overline{B}{}^0_s$ system.  
Consideration of the sample sizes suggests that it will be possible to measure the relevant observables with similar or better precision to previous experiments, as reported in Ref.~\cite{Abada:2019lih}.  A particular strength of the FCC-ee flavour-programme will be the ability to make very sensitive studies of modes containing neutrals, with much larger sample sizes than will be available at Belle II.   This possibility will enable FCC-ee to harness a very wide range of charm-meson decay modes in measurements of $C\!P$ asymmetries in $B^- \to DK^-$ (where $D$ indicates a superposition of $D^0$ and $\bar{D}^0$ and $B_s \to D_s^{(*)\pm} K^\mp$, which are sensitive to the angle $\gamma$. It is expected that the flavour-tagging efficiency will be significantly higher than at the LHC, bringing corresponding gains for time-dependent measurements.
Other exciting possibilities include modes relevant for the angle $\alpha$;  for example, precise measurements of the time-dependent $C\!P$ asymmetries in $B^0 \to \pi^0\pi^0$ can be performed making use of Dalitz decays. Around 2.5 thousand  $B^0 \to \pi^0 (\to \gamma \gamma) \pi^0 (\to e^+ e^- \gamma) $ events can be expected,  assuming electron and $\pi^0 \to \gamma \gamma$ reconstruction efficiencies similar to those achieved with the LEP detectors.  

\begin{figure}
\centering
    \begin{subfigure}[b]{.45\textwidth}
    \resizebox{\textwidth}{!}{\includegraphics{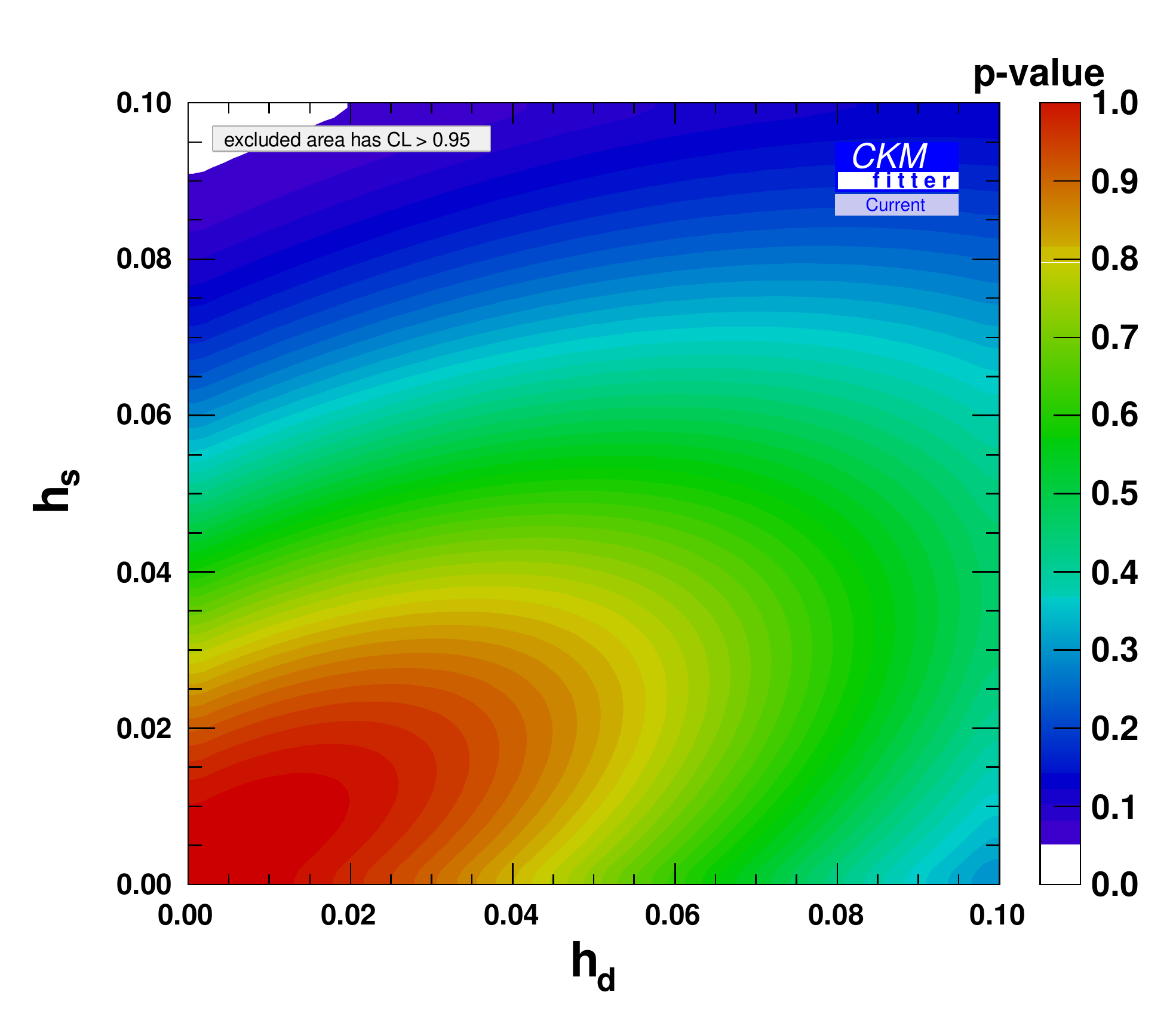}}
    \end{subfigure}
    \begin{subfigure}[b]{.45\textwidth}
    \resizebox{\textwidth}{!}{\includegraphics{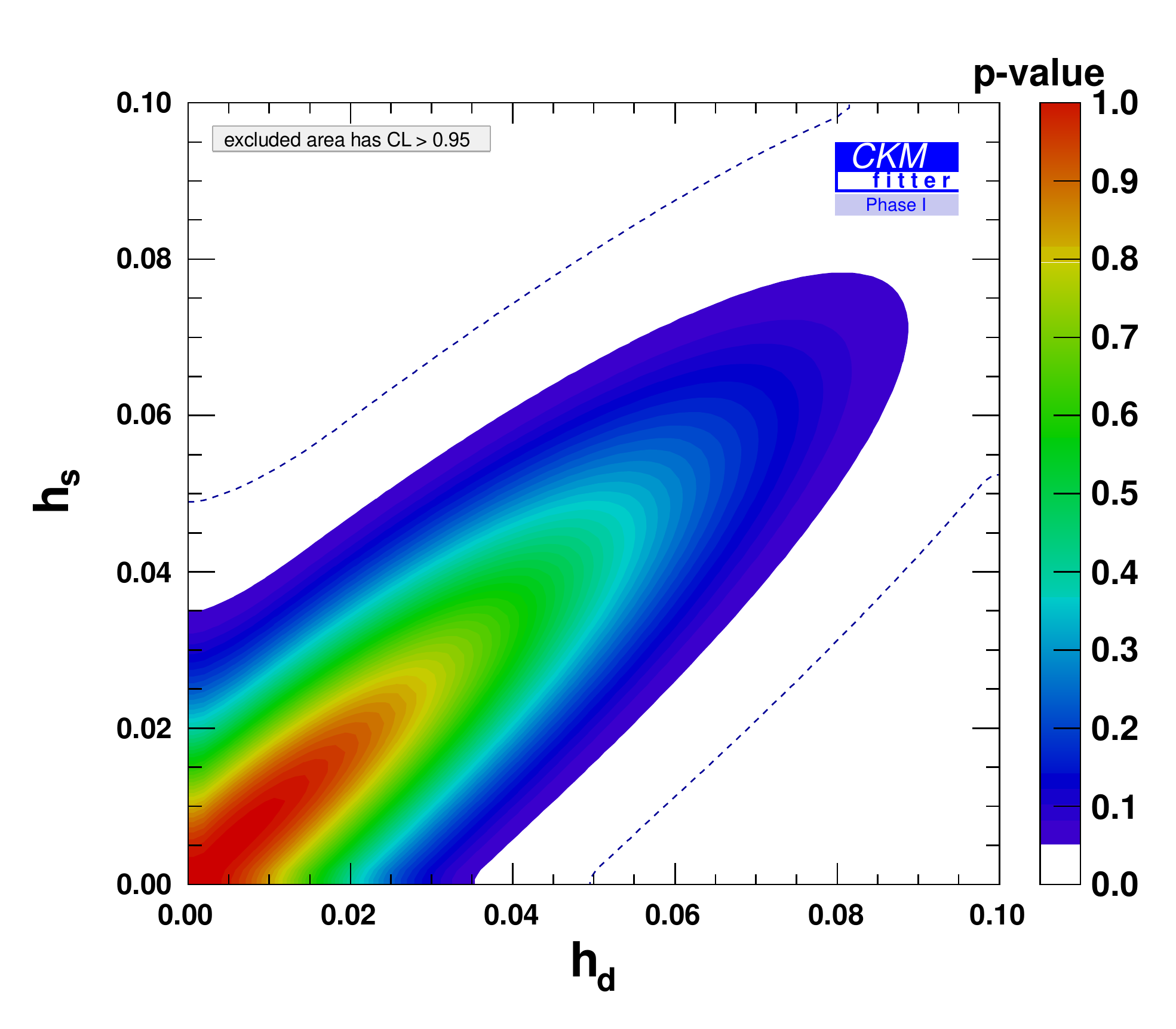}}
    \end{subfigure}
    \vskip\baselineskip
    \begin{subfigure}[b]{.45\textwidth}
    \resizebox{\textwidth}{!}{\includegraphics{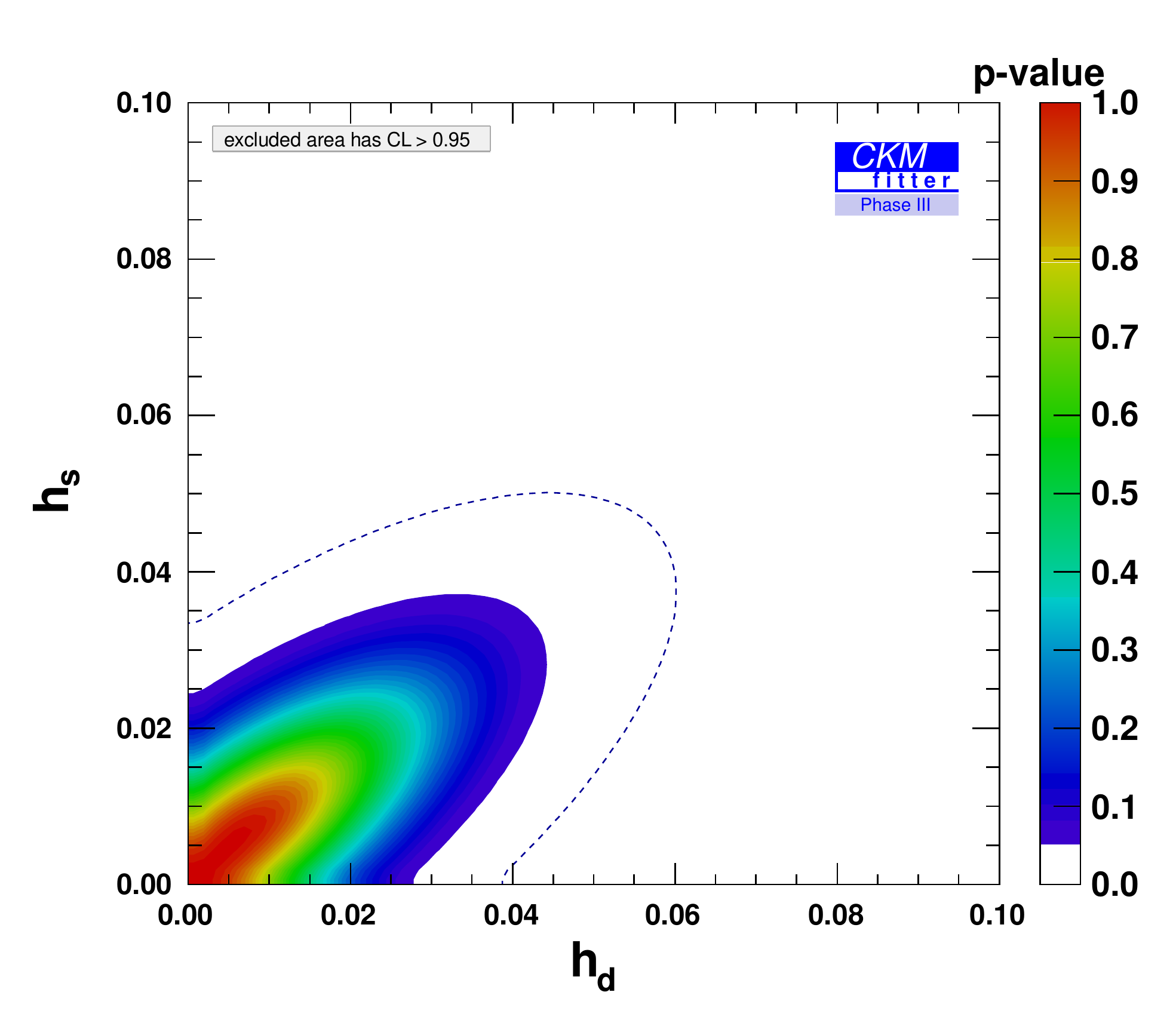}}
    \end{subfigure}
\caption{
Sensitivity to $h_d-h_s$ parameters in $B_{d}$ and $B_{s}$ mixing
with current sensitivity (top left), the anticipated constraints after LHCb Upgrade I and Belle II operation (top right), and the FCC-ee expectation (bottom). The dotted curves show the 99.7\%~C.L. ($3\sigma$) contours.  All plots are made with the SM inputs $h_d=h_s=0.$   Taken from Ref.~\cite{Charles:2020dfl}.
}
\label{fig:hdhs}
\end{figure}

It is quite possible, however, that the most significant impact of FCC-ee in this domain will come from the measurement of other observables.  In making this argument, it is useful to recall the results of a model-independent study of possible BSM contributions to the $B^0-\overline{B}{}^0$ and $B^0_s-\overline{B}{}^0_s$ mixing amplitudes, reported in Ref.~\cite{Charles:2020dfl}.  The box diagrams that drive the oscillations and carry $C\!P$-violating phases are natural entry points for any heavy degrees-of-freedoms from BSM physics.  The mixing amplitudes can be modelled with complex numbers of modulus $h_q$ ($q=d,s$), each multiplying the SM $B^0_q$  mixing Hamiltonian matrix element. 
Figure~\ref{fig:hdhs} shows the constraints on these BSM parameters (assuming no BSM signal) for three scenarios: current measurements, the foreseen  LHCb upgrade and Belle II experiments measurements, and at FCC-ee.
The projected performance at Belle II and LHCb are based on the numbers reported in Refs.~\cite{lhcbupgrade2,Kou_2019}, and reasonable assumptions concerning progress in lattice QCD~\cite{Cerri:2018ypt}. 
The FCC-ee inputs include extrapolations of the current $\beta$ and $\gamma$ angle measurements to the expected statistics at the $Z$ pole~\cite{Abada:2019lih}. As no dedicated studies have yet been performed, we conservatively assume that the measurements of $|V_{ub}|$ and the mixing frequencies, as well as knowledge of hadronic parameters, will remain unchanged from the end of the HL-LHC era~\cite{Cerri:2018ypt,Kou_2019}. 
The FCC-ee constraints benefit in addition from precise determinations of three key parameters: $|V_{cb}|$, and the semileptonic asymmetries $a^d_{\rm sl}$ and $a^s_{\rm sl}$.

\begin{itemize}
    \item The sides of the Unitarity Triangle are normalised by the CKM element $V_{cb}$. In Fig.~\ref{fig:hdhs} the sensitivity to the BSM parameters, even with the very large sample sizes of LHCb Upgrade I and Belle II accumulated by around 2030, start to become limited by the knowledge of this parameter, which comes from measurements of semileptonic $B$ decays.  These measurements can be made both with inclusive and exclusive decays, but in both cases require hadronic input, {\it e.g.} from lattice QCD, for $|V_{cb}|$ to be extracted.
    At FCC-ee another approach will open up that has no such systematic limitation.
    Several $10^8$ $W$ boson decays will be collected when operating FCC-ee at the $W^+W^-$ threshold and above.  By benefitting from the excellent vertexing capabilities that will be available at FCC-ee, and selecting events with $b$-tagged and $c$-tagged jets ~\cite{Behnke:2013xla,FCC_Vcb}, it may be possible to improve the knowledge of $|V_{cb}|$ up to an order of magnitude with respect to current knowledge.
    It will also be possible to make very precise measurements of the magnitude of other CKM elements, {\it e.g.} $|V_{cs}|$, using on-shell $W$ decays. Similar studies, albeit with lower statistical precision, will be possible with the large $t\bar{t}$ sample that will be accumulated  (see Ref.~\cite{Harrison:2018bqi} for a proposal first made in the context of the LHC).  
    
    \item The semileptonic asymmetries 
    \begin{equation}
        a^q_{\rm sl} \equiv \frac{\Gamma(\overline{B}{}^0_q \to \bar{f}) - \Gamma({B^0_q} \to f)}{\Gamma(\overline{B}{}^0_q \to \bar{f}) + \Gamma({B^0_q} \to f)},
    \end{equation}
    which are so-called as semileptonic decays are typically chosen for the flavour-specific final-state $f$, probe the $C\!P$-violating phases in $B^0_q - \bar{B}{}^0_q$ oscillations ($q=d,s$).  In the SM these asymmetries are very small, but precisely predicted: $a^d_{sl} = (-4.73 \pm 0.42) \times 10^{-4}$ and $a^s_{sl} = (2.06 \pm 0.18) \times 10^{-5}$~\cite{Lenz:2019lvd}.  Hence measurements of $a^d_{\rm sl}$ and $a^s_{\rm sl}$ are very valuable in providing sensitivity to  $h_d$ and $h_s$. LHCb Upgrade II will attain a statistical sensitivity of a few $10^{-4}$ for both observables, but controlling systematic effects from production and detection asymmetries at the same level of precision will be very challenging~\cite{lhcbupgrade2}. A solenoidal detector at FCC-ee will have clear advantages in this regard, and can reach an overall sensitivity of a few $10^{-5}$~\cite{FCC_aSL}.   
    
\end{itemize}

In addition to the measurement of CKM-related observables, FCC-ee will perform studies of a wide range of suppressed flavour-changing-neutral-current (FCNC) processes, such as dileptonic and semileptonic modes that are driven by electroweak penguins and box diagrams in the SM, for example $B^0 \to \mu^+\mu^-$, $B^0_s \to \mu^+\mu^-$, $B^0_s \to \tau^+ \tau^-$, $b \to s(d) \ell^+\ell^-$, where $\ell$ represents either an electron or a muon, $b \to s(d) \tau^+ \tau^-$ and  $b \to s(d) \nu \overline{\nu}$. This programme will be extended to the analysis of favoured, but experimentally challenging, modes, where the SM predictions are reliable and BSM effects could be pronounced, for example the decays $B_c^+ \to \mu^+\nu$ and $B_c^+ \to \tau^+\nu$. Promising exploratory studies have been performed for the latter mode at a $Z$ factory~\cite{Zheng:2021xuq,Amhis:2021cfy}.
The analysis of these channels, together with that of radiative FCNCs in both the beauty and charm sectors, will allow for stringent tests of the SM and have high discovery potential in probing for new-physics extensions.  FCC-ee will be able to complement many of the studies performed at Belle II and LHCb, and significantly extend the measurement programme in several critical areas.  

\begin{figure}
    \centering
    \resizebox{0.6\textwidth}{!}{\includegraphics{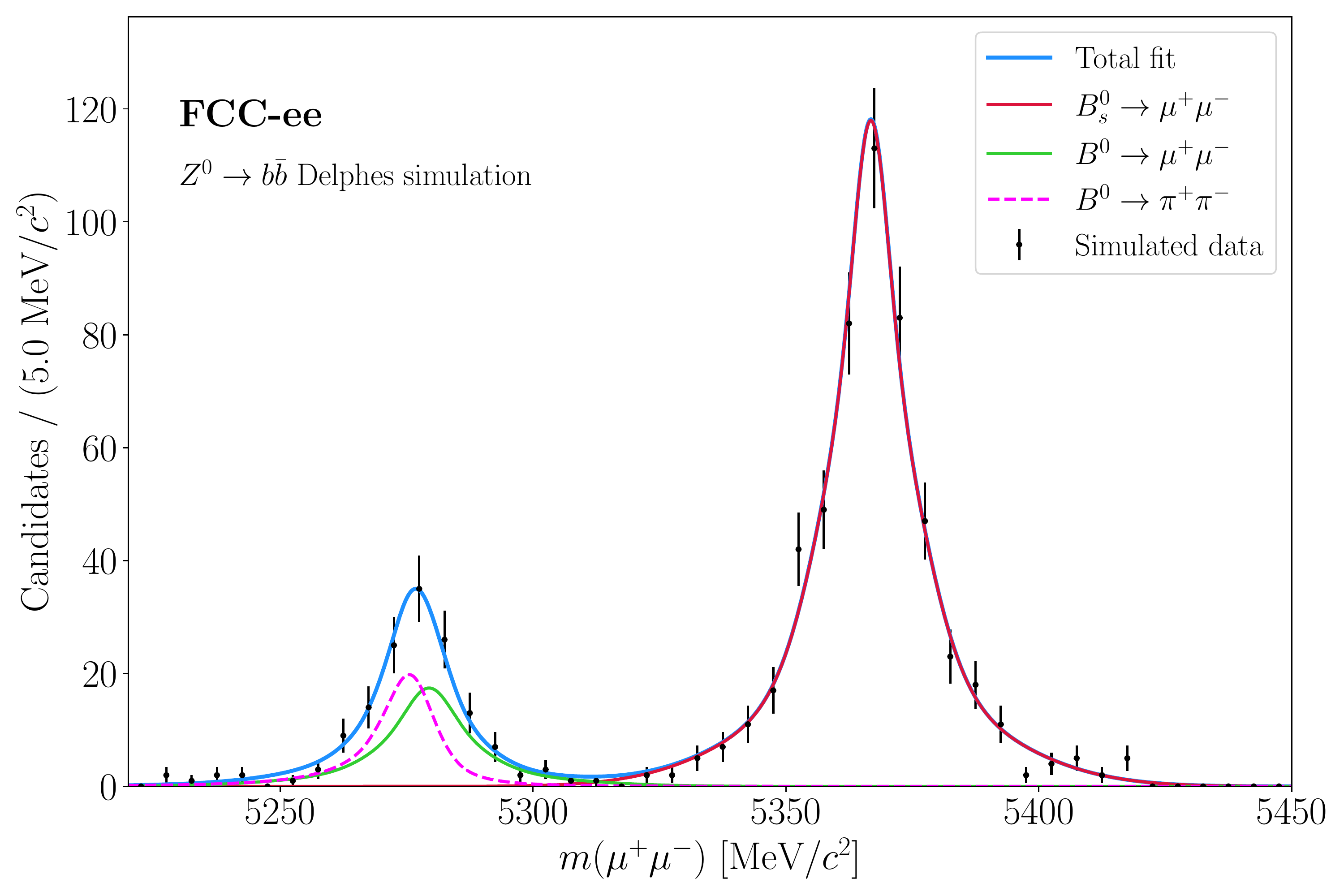}}
    \caption{Reconstructed invariant mass of $B^0 \to \mu^+\mu^-$ and $B^0_s \to \mu^+\mu^-$ signals for $5 \times 10^{12}$ $Z^0$ decays.  Also shown is the background contribution from misidentified $B^0 \to \pi^+\pi^-$ events~\cite{Hill}.}
    \label{fig:btomumu}
 \end{figure}

To illustrate these possibilities more clearly, we here discuss the `flagship' ultra-rare channels $B^0_s \to \mu^+\mu^-$ and $B^0 \to \mu^+\mu^-$.  In Runs 1 and 2 at the LHC the focus has been on measuring the branching fraction of $B^0_s \to \mu^+\mu^-$~\cite{Aaij:2017vad,Aaboud:2018mst,Sirunyan:2019xdu} and comparing with the precise predictions of the SM~\cite{Beneke:2019slt}.  In coming years, and at FCC-ee, emphasis will shift to  the search for the even rarer sister mode, $B^0 \to \mu^+\mu^-$, and the measurement of the ratio of the $B^0 \to \mu^+\mu^-$ to $B^0_s \to \mu^+\mu^-$ branching fractions, which is a powerful test of minimal flavour violation.  Further important observables will include the $B^0_s \to \mu^+\mu^-$ effective lifetime~\cite{DeBruyn:2012wk}, where proof-of-principle measurements already exist~\cite{Aaij:2017vad}, and measurements of $C\!P$ asymmetries, which currently are not feasible.  In all these studies FCC-ee will excel, with particular benefits coming from the excellent mass resolution, which will clearly separate the $B^0$ and $B^0_s$ signals, high-performance flavour tagging (necessary for the $C\!P$-asymmetry measurements), and low backgrounds.  Figure~\ref{fig:btomumu} shows the signal peaks that are expected with $5 \times 10^{12}$ $Z^0$ decays and reasonable assumptions on the detector performance.   
These peaks contain around 540 and 70 $B^0_s \to \mu^+\mu^-$ and $B^0 \to \mu^+\mu^-$ decays, respectively~\cite{Hill}.
Also included is the background arising from $B^0 \to \pi^+\pi^-$ decays,  assuming a double $\pi \to \mu$ misidentification probability of $2 \times 10^{-5}$, which lies under the $B^0 \to \mu^+\mu^-$ signal. Controlling this source of contamination will be an important consideration in the detector optimisation.

\begin{figure}
    \centering
    \resizebox{0.6\textwidth}{!}{\includegraphics{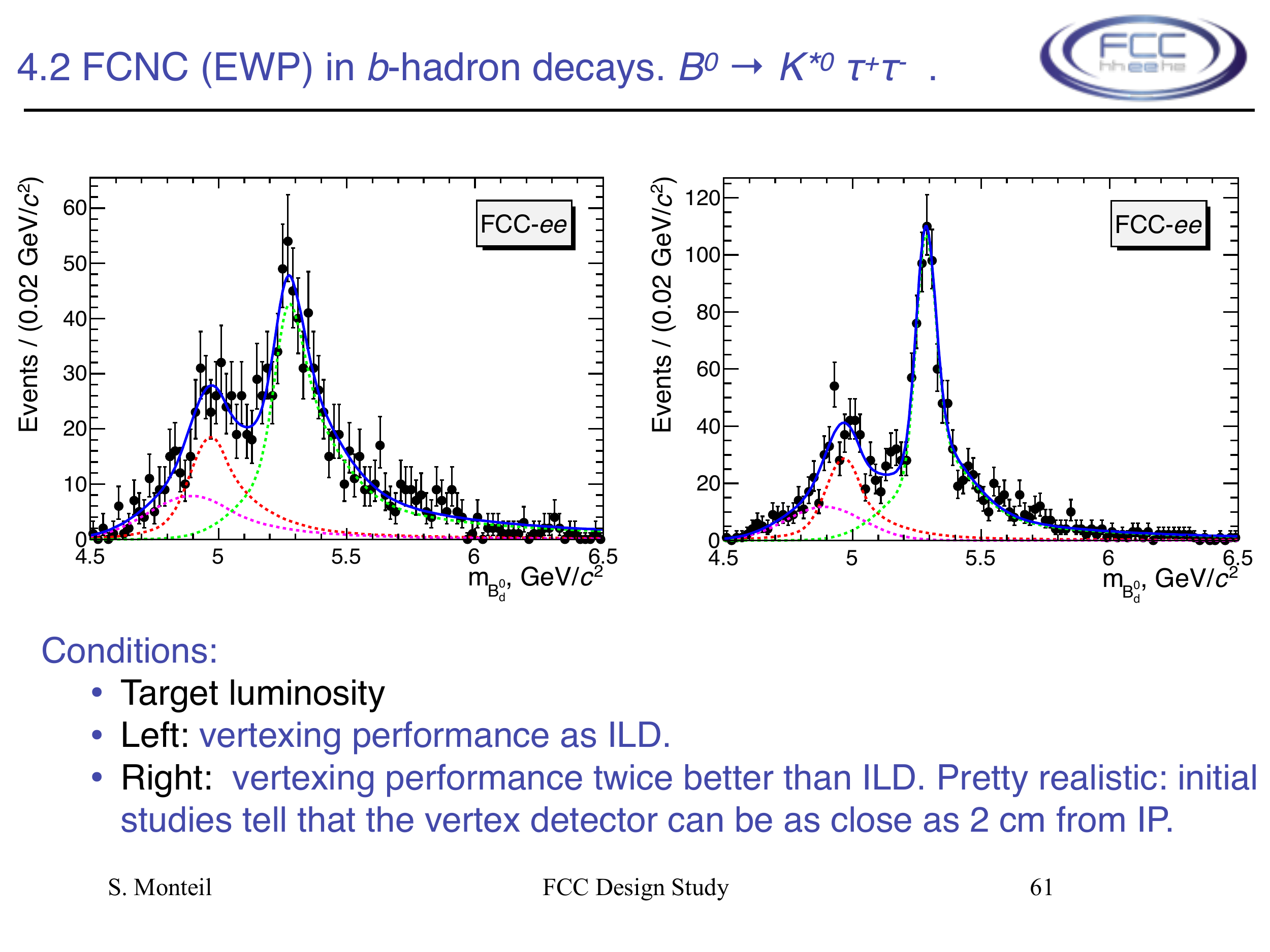}}
    \caption{Reconstructed invariant-mass of $B^0 \to K^*(892) \tau^+\tau^-$ candidates determined by using constraints on the production and decay vertices. The assumed resolutions on the primary and secondary vertices cartesian coordinates  are 3 and 7 ${\rm\mu}$m, respectively. Two background sources are considered: (red) $B^0\to D_s^-\overline{K}{}^{*0}(892)\tau^+ \nu_{\tau}$ and (pink) $\overline{B}_s \to D_s^-D_s^+\overline{K}{}^{*0}(892)$, with the relative branching fractions appropriately scaled to their estimated SM values.}
    \label{fig:Bd2Kstartautau}
 \end{figure}

Final states involving $\tau$ leptons also warrant further discussion on account of the unique opportunities that will exist at FCC-ee~\cite{Altmannshofer:2014cfa,Varzielas:2015iva,Bordone:2018nbg}. 
 Studies reported in Refs.~\cite{Kamenik:2017ghi,Li:2020bvr} have established 
a proof-of-principle of the kinematical fits that determine the missing neutrino missing momentum for three-prongs $\tau$ decays, as described in a  companion essay in this volume.  This approach requires the detector to be highly hermetic and have excellent resolution for reconstructing the production and decay vertices.    Figure~\ref{fig:Bd2Kstartautau} illustrates these capabilities for the mode $B^0 \to K^*(892) \tau^+\tau^-$, showing the reconstructed mass peak in a fast simulation with a parametric detector response, and including the contribution of a subset of background decay modes sharing the very same signal final state. This signal channel is of great interest, particularly in light of the hints of the violation of lepton universality seen in the companion decays involving the first two generations, but cannot be well studied at Belle II or LHCb Upgrade II unless its branching fraction is greatly enhanced with respect to the SM expectations. 
Another example is the decay channel $B_c^+ \to \tau^+\nu$, where a recent analysis, already mentioned, has demonstrated good prospects at FCC-ee~\cite{Amhis:2019ckw}.
These promising initial studies must be refined, with all background sources comprehensively accounted for, and the detector design optimised to maximise performance for this very important category of decay channel.

The brief survey of interesting observables does not exhaust the list of $b$-physics opportunities at FCC-ee, but highlights the unique capabilities of the facility.  It is clear from the examples given that many important measurements can be made that will complement well the strengths of the LHC flavour programme. The ability to exploit the full range of $b$-hadrons, and with a high Lorentz boost, gives FCC-ee an advantage over Belle II that is even more important than its statistical superiority.  Furthermore, there will also be similarly important and exciting studies to perform in the domains of charm physics and heavy-flavour spectroscopy.

\input{det_requirements}


\section{Conclusions}
\label{section:conclusion}

In this essay we have discussed the features of the $Z^0$ factory that make it a very promising environment for heavy-flavour physics, combining most of he advantages of Belle II and LHCb experiments, and have attempted to foresee the  experimental and theoretical landscape of flavour physics at the dawn of the FCC-ee.  We then reviewed selected key observables, where the impact of FCC-ee will be most pronounced, and which are promising in probing for BSM effects.
These measurements place specific demands on the detector design, particularly in the areas of vertexing, calorimetery and particle identification.  These requirements are not necessarily the same as those required for electroweak and Higgs physics, and motivate a machine with four interaction points.  The outstanding physics opportunities and stimulating detector challenges make the heavy-flavour programme one of the most exciting and attractive prospects at FCC-ee.

\section*{Acknowledgements}
We thank Donal Hill for helpful discussions and contributing material to this essay.  

%
%
%

\bibliographystyle{jhep}
\bibliography{references}
\end{document}

%% file: introduction.tex
\section{Introduction: FCC-ee as a flavour factory}
\label{section:intro}

Measurements in flavour physics have played a central role in the construction of the Standard Model (SM).
Studies in the kaon sector led to the GIM mechanism and the prediction of the charm quark~\cite{GIM}; the observation of $C\!P$ violation suggested the existence of a third generation~\cite{Kobayashi:1973fv},  and the first measurements of $B^0-\overline{B}{}^0$  oscillations immediately indicated the top quark to be very massive~\cite{ALBRECHT1987245}.
Now complete, the SM poses many questions that are intimately connected to flavour.
Why are there three generations? What explains the hierarchy in quark masses and the distinctive structure of the Cabibbo-Kobayashi-Maskawa matrix?  What is the nature of the non-SM contributions to $C\!P$ violation that are required to explain the matter-antimatter asymmetry of the universe?  
Furthermore, flavour physics is known to be a powerful tool of discovery, as the contribution of loop amplitudes to many of the processes of interest provide natural entry points for new, massive particles.  Indeed, the broad consistency of flavour observables with SM predictions was the first warning sign that TeV-scale non-SM physics (hence referred to a `new physics') would not be observed at the LHC.  For these reasons flavour physics,  in particular studies of beauty, charm, and of $\tau$ leptons, is a vibrant field of study, with the current flagship experiments being LHCb at the Large Hadron Collider,  and Belle II operating in the $e^+e^-$ environment at the $\Upsilon(4S)$.

\begin{table}
    \caption{Advantageous attributes for flavour-physics studies at Belle II ($\Upsilon(4S)$), the LHC ($pp$) and FCC-ee ($Z^0$).}
    \centering
    \begin{tabular}{lccc} \hline
    Attribute & $\Upsilon(4S)$ & $pp$ & $Z^0$ \\ \hline
All hadron species &  & \cmark & \cmark \\
High boost         &  & \cmark & \cmark \\
Enormous production cross-section & & \cmark &  \\
Negligible trigger losses & \cmark &  &  \cmark \\
Low backgrounds & \cmark &  &  \cmark \\
Initial energy constraint & \cmark &  & (\cmark)  \\ \hline
    \end{tabular}
    \label{tab:attributes}
\end{table}

Table~\ref{tab:attributes} compares the advantages for flavour-physics studies at an $e^+e^- \to \Upsilon(4S) \to b\bar{b}$ experiment, such as Belle~II, a $pp \to b\bar{b} X$ experiment, such as LHCb, and an experiment that relies on $e^+e^- \to Z^0 \to b\bar{b}$ production, such as would be the case at FCC-ee. 
It can be seen that the $Z^0$ environment combines most of the advantages of Belle~II and LHCb.  
For the former these are the high signal-to-noise and fully efficient trigger, as well as a very high geometrical acceptance; for the latter they are the production of the full spectrum of hadrons, and the high boost.
The momenta of $b$ and $c$ hadrons produced at the $Z^0$ are not known a priori, in contrast to the $\Upsilon(4S)$, although their distribution is very well understood.  The momentum of the produced tau leptons is of course perfectly known in both $e^+e^-$ environments.

The one disadvantage that the $Z^0$ has in comparison with the LHC is the production cross section,  but this is partially mitigated at FCC-ee by the enormous luminosity that is foreseen.  Table~\ref{tab:flavouryields} gives the yields for each $b$-hadron species that will be produced in $5 \times 10^{12}$ $Z^0$ decays~\footnote{Note that about a factor of two more $Z^0$ decays can be recorded if the design of the FCC-ee evolves towards a four interaction-points layout.}.    The number of $b\bar{b}$ pairs from which these yields arise is around fifteen times larger than that expected at Belle~II.  As will be explained below, the particular advantages of the $Z^0$ environment will allow for many studies that are complementary or more sensitive to those foreseen at LHCb and its upgrades.  There will also be great opportunities in charm and tau physics, for which yields are also listed in Table~\ref{tab:flavouryields}.  In tau physics, in particular, the FCC-ee will have unsurpassed physics reach in almost all measurements, as is discussed in companion essays in this volume.

\begin{table}
\caption{Yields of heavy-flavoured particles produced at FCC-ee for $5 \times 10^{12}$ $Z^0$ decays. The charge conjugate states have the same production yields.  These yields are computed using the Z branching fractions and hadronisation rate reported in Refs.~\cite{Amhis:2019ckw,Zyla:2020zbs}. The $B_c$ hadronisation fraction is assumed to be $f_{B_c} = 2 \times 10^{-3}$~\cite{Aaij:2019ths}.\vspace*{0.1cm}} 
\centering
\begin{tabular}{cccccccc} 
 \hline
 Particle species & $B^0$ & $B^+$& $B^0_s$ & $\Lambda_b$ & $B_c^+$ & $c \overline{c}$ & $\tau^-\tau^+$ \\
   \hline

      Yield ($\times 10^9$)    & $310$ &  $310$ &  $75$  & $65$ & $1.5$ & $600$ & $170$ \\
  \hline
\end{tabular} 
\label{tab:flavouryields}
\end{table}

We also note that the proposed running strategy of FCC-ee, with the intention to collect data at several collision energies, will open up flavour possibilities that are not restricted to the $Z$ pole.  The decays of on-shell $W$ bosons will provide a particularly rich laboratory for studies of the CKM matrix, as is described in more detail below.

In this brief report we first survey the likely landscape in quark-flavour physics in the mid to late 2030s, for both theory and experiment.  We then outline the opportunities that will exist in this domain at FCC-ee, highlighting certain key measurements.  We conclude  by presenting the key detector characteristics that an experiment should possess in order to realise these opportunities.

%% file: landscape.tex
\section{Quark-flavour physics at the dawn of FCC-ee: the theoretical and experimental landscape}
\label{section:physics}

The sensitivity to virtual particles and phases that is afforded by flavour observables give them excellent sensitivity to new physics at mass scales far in excess of those achievable in direct searches.  In this sense the goals of flavour-physics studies at FCC-ee are similar and complementary to those directed at electroweak precision observables. Already some intriguing hints of new physics effects have been seen in possible violations of lepton universality in both penguin-mediated processes and in tree decays.  Rather than speculate about the evolution of these anomalies, we here choose to make more general remarks on the status of the flavour-measurement programme, with a focus on where further experimental progress will be valuable during the FCC-ee era.

\begin{figure}
    \centering
    \resizebox{0.8\textwidth}{!}{\includegraphics{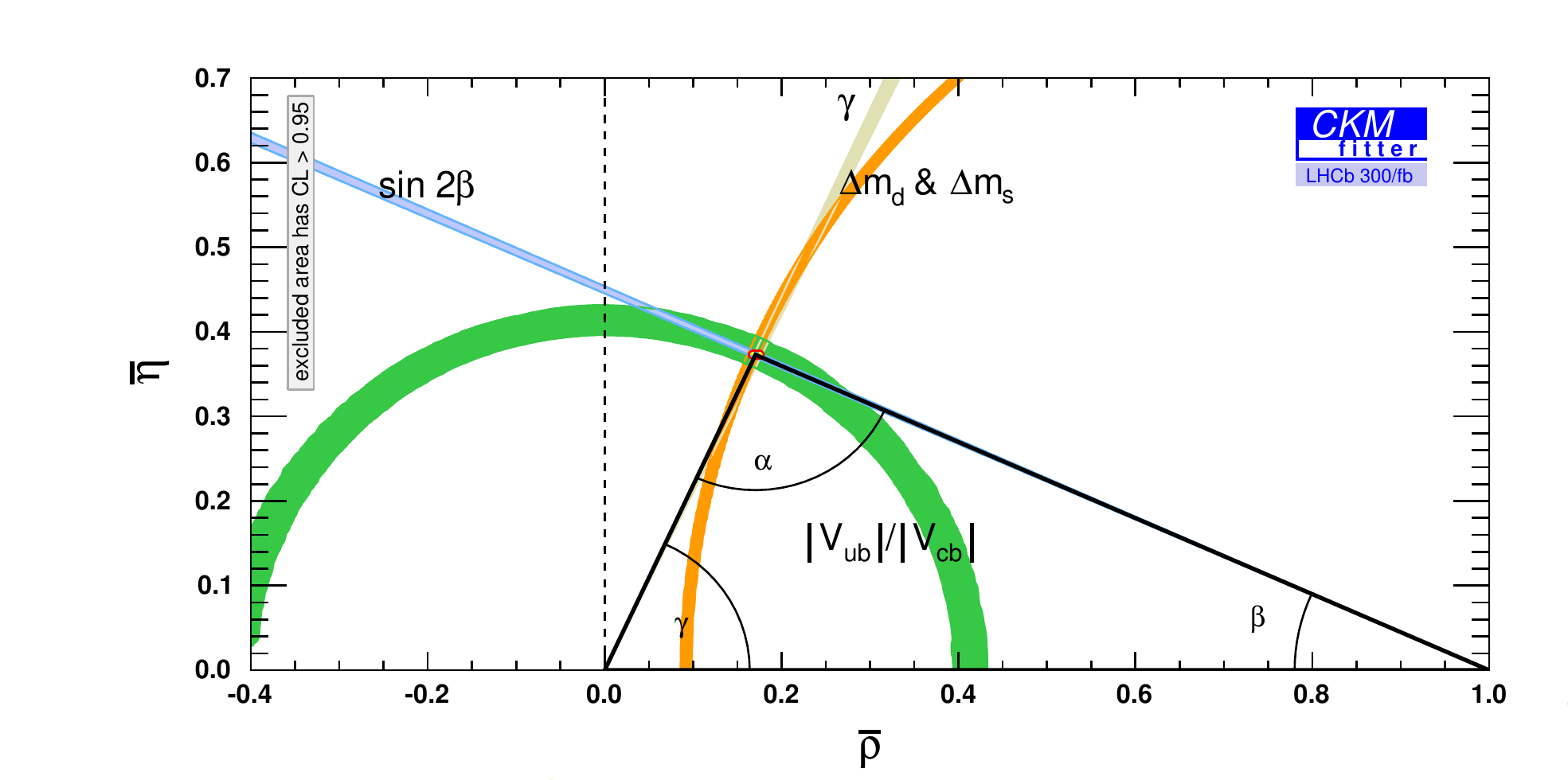}}
    \caption{Possible status of the Unitarity Triangle in the late 2030s, assuming LHCb measurements alone and improvements in lattice QCD~\cite{lhcbupgrade2}.}
    \label{fig:lhcbup2}
 \end{figure}

According to current planning, the 2030s will see operation of the second upgrade of LHCb, which will accumulate around 300\,fb$^{-1}$ in integrated luminosity, and collect sample sizes in excess of 50 times larger than those so far analysed~\cite{lhcbupgrade2}. ATLAS and CMS will continue to contribute strongly to the LHC flavour programme in selected areas, for example the measurement of the weak phase $\phi_s$ and studies of $B^0 \to \mu^+\mu^-$ and  $B^0_s \to \mu^+\mu^-$. Belle~II will integrate around 50\,ab$^{-1}$, with the prospect of further running now under active discussion. In parallel to the great steps forward in experimental knowledge these projects will bring, there are expected to be corresponding advances in lattice QCD calculations, and the development of other theoretical tools.  Figure~\ref{fig:lhcbup2} shows the possible status of the Unitarity Triangle after LHCb Upgrade~II operation, based on results from that experiment alone, and assuming plausible lattice QCD improvements.  It makes the pessimistic assumption that all results will be in agreement.  Similarly high precision will be achieved in studies of `rare decays', {\it i.e.} suppressed flavour-changing neutral-current (FCNC) processes.

 Many of the observables in $C\!P$-violation studies, and others contributing to Unitarity Triangle tests, are theoretically pristine, and so warrant continued experimental attention throughout the coming decades.  The angle $\gamma$, for example, may be measured in $B^- \to DK^-$ decays with a relative theoretical uncertainty of $10^{-6}$ or better~\cite{Zupan:2011mn}.  Other measurements, that may become limited by systematic biases at the LHC, will benefit from the cleaner and very different environment of the FCC-ee. Examples include studies of semileptonic $C\!P$-violating asymmetries, and determinations of $|V_{ub}/V_{cb}|$ performed with $B^0_s$ mesons and $\Lambda_b$ baryons that are not accessible at Belle~II. Here the hadronic systematics may be different to those that apply for measurements with $B^{0,\pm}$ mesons, and the spin of the baryon brings potentially useful new information.  Very importantly, there will remain much to be learned from $b$-hadron decays involving $\tau$ leptons, which are very challenging at LHCb, and where the $B$-factory studies will be limited by sample size, as well as being restricted to $B^0$ and $B^-$ mesons.

%% file: det_requirements.tex
\section{Detector requirements: towards a dedicated heavy-flavour experiment}
\label{section:detector}

Although there are commonalities between the experimental requirements of all physics goals at FCC-ee, for example the need to cope with collimated jets of particles, 
a detector optimised for beauty, charm and, indeed, tau physics will have capabilities that are rather different to those that are necessary for, say, Higgs studies. The experimental layouts considered until now at FCC-ee have, understandably, been designed with the higher-energy requirements in mind. There are scenarios, however, under which an experiment largely or wholly optimised for heavy-flavour physics  is conceivable.  For example, if the machine design evolves to a four-interaction point collider then there will be a strong argument for diversity in the attributes of the experiments that are deployed. 
Here we sketch out the essential features of a heavy-flavour experiment, paying particular attention to the demands on vertexing, calorimetry and hadron particle identification (PID).

Precise vertexing is essential for all heavy-flavour measurements, for example playing a particularly important role in time-dependent $B^0_s$ studies where the proper-time uncertainty must be small compared to the $\sim$350\,fs oscillation period of the $B^0_s$ meson in order not to dilute the $C\!P$ asymmetries.  The performances of the vertex detectors proposed for the IDEA and CLD detectors are probably adequate in this regard, with resolutions of a few microns for multi-track vertices~\cite{Abada:2019zxq}.  However, for other studies there will be significant gains if even better performance can be attained.  A case in point is the study of $b$-hadron decays involving tau leptons, discussed above. Here the partial reconstruction of the decay relies on a kinematic fit that takes as input the tau decay vertex and, if available, that of the $b$ hadron.  Current detector designs aim for a resolution of around 5\,{$\mu$m} for a three-track vertex, which will limit the effective invariant-mass resolution for this class of decays. Starting with the ALPIDE-style sensor design~\cite{AGLIERIRINELLA2017583,Snoeys:2017hjn} already proposed for the IDEA experiment, various improvements can be envisaged that would lead to reduced vertex uncertainties.  These include a smaller beam-pipe radius, the deployment of curved sensors, improved hit resolution and the implementation of air cooling.

 The relatively low-multiplicity environment of FCC-ee presents an outstanding opportunity to make high precision studies of heavy-flavour decays involving photons, $\pi^0$ and $\eta$ mesons.  (Although important results in this domain have emerged from LHCb, the situation there is a priori very challenging, due to the high level of combinatorics).  Achieving this goal requires excellent energy resolution in order to obtain an invariant-mass resolution that is comparable to that obtainable for decays involving charged tracks only.  This attribute will be even more critical at FCC-ee than at the $B$-factories, as it will be necessary to separate $B^0$ and $B^0_s$ decays to the same final state. To pick but one example, without excellent mass resolution, $b \to d \gamma$ penguin transitions in $B^0_s$ decays will be overwhelmed by $B^0$ decays involving the much more prevalent $b \to s \gamma$ process. It will be necessary to maintain this level of performance down to low momenta.  For example, an important class of $C\!P$-violation measurements requires separation of the decays $D^{\ast 0} \to D^0 \pi^0$ and $D^{\ast 0} \to D^0 \gamma$~\cite{Bondar:2004bi}.  The isolation criteria that must be imposed in selecting decays such as $B_c^+ \to \tau^+ \nu_{\tau}$, discussed above, point to similar requirements.  Meeting these physics goals mandates a crystal-based electromagnetic calorimeter, as appreciated by CLEO and the $B$-factories.
 To set the scale, the BaBar detector was equipped with a thallium doped caesium-iodide calorimeter that achieved an energy resolution with a stochastic term better than 3\%~\cite{TheBABAR:2013jta}.
 In contrast, hadron calorimetry is not an essential capability for flavour measurements, but consideration should be given to $K^0_L$ detection, which at BaBar was performed by combining information from the ECAL and the instrumented flux return~\cite{Aubert:2009aw}.
 
 An extensive flavour-physics programme requires high-performance PID over a wide momentum range.  A powerful component of the flavour tagging, necessary for time-dependent
 $C\!P$-violation measurements, comes from noting the charge of kaons in the event not associated with the signal decay. These kaons are typically below 10\,GeV/$c$.  In contrast, kaons from the signal decay are harder, with some modes giving rise to spectra extending  beyond 30\,GeV/$c$.  This wide span of momentum necessitates different solutions than those that were adequate at the $B$-factories, with tighter space constraints than exist for LHCb.    The cluster-counting of ionisation deposits that is intended for the IDEA drift chamber would largely meet these requirements, if it performs to specification~\cite{Abada:2019zxq}.  Other technologies, such as time-of-flight measurements using a TORCH detector~\cite{CHARLES2011173,BHASIN2020163671}, would provide useful performance, but not at high momentum.  A RICH system with more than one radiator could encompass the necessary momentum range, but would unavoidably take up ${\cal{O}}(1\,{\rm m})$ of radial space in the detector, which would then have consequences for the tracking volume and also degrade the energy resolution, if placed in front of the ECAL.  
 Solving this problem is an interesting challenge for the overall detector optimisation.
 More extensive discussion on this topic may be found in a companion essay in this volume.